\documentclass[pra,twocolumn,preprintnumbers,amsmath,amssymb,superscriptaddress]{revtex4}

\usepackage{color}
\usepackage{graphicx}
\usepackage{dcolumn}
\usepackage{bm}
\usepackage{hyperref}
\usepackage{enumitem}
\usepackage{balance}
\usepackage{comment}
\usepackage{cleveref}

\begin{document}


\title{Remote state preparation of two-component Bose-Einstein condensates}

\author{Manish Chaudhary}
\affiliation{State Key Laboratory of Precision Spectroscopy, School of Physical and Material Sciences, East China Normal University, Shanghai 200062, China}
\affiliation{New York University Shanghai, 1555 Century Ave, Pudong, Shanghai 200122, China}

\author{Matteo Fadel}
\affiliation{Department of Physics, University of Basel, Klingelbergstrasse 82, 4056 Basel, Switzerland}

\author{Ebubechukwu O. Ilo-Okeke}
\affiliation{New York University Shanghai, 1555 Century Ave, Pudong, Shanghai 200122, China}
\affiliation{Department of Physics, School of Physical Sciences, Federal University of Technology, P. M. B. 1526, Owerri, Imo State 460001, Nigeria}

\author{Alexey N. Pyrkov}
\affiliation{Institute of Problems of Chemical Physics of Russian Academy of Sciences, Acad. Semenov av. 1, Chernogolovka, Moscow Region, Russia 142432}

\author{Valentin Ivannikov}
\affiliation{New York University Shanghai, 1555 Century Ave, Pudong, Shanghai 200122, China}
\affiliation{State Key Laboratory of Precision Spectroscopy, School of Physical and Material Sciences, East China Normal University, Shanghai 200062, China}
\affiliation{NYU-ECNU Institute of Physics at NYU Shanghai, 3663 Zhongshan Road North, Shanghai 200062, China}

\author{Tim Byrnes}
\affiliation{New York University Shanghai, 1555 Century Ave, Pudong, Shanghai 200122, China}
\affiliation{State Key Laboratory of Precision Spectroscopy, School of Physical and Material Sciences, East China Normal University, Shanghai 200062, China}
\affiliation{NYU-ECNU Institute of Physics at NYU Shanghai, 3663 Zhongshan Road North, Shanghai 200062, China}
\affiliation{National Institute of Informatics, 2-1-2 Hitotsubashi, Chiyoda-ku, Tokyo 101-8430, Japan}
\affiliation{Department of Physics, New York University, New York, NY 10003, USA}

\date{\today}

\begin{abstract}
A protocol for remote state preparation is proposed for spin ensembles, where the aim is to prepare a state with a given set of spin expectation values on a remote spin ensemble using entanglement, local spin rotations, and measurements in the Fock basis.  The spin ensembles could be realized by thermal atomic ensembles or spinor Bose-Einstein condensates. The protocol works beyond the Holstein-Primakoff approximation, such that spin expectation values for the full Bloch sphere can be prepared.  The main practical obstacle is the preparation of the maximally entangled state between the spin ensembles.  To overcome this, we examine the protocol using states based on the two-axis two-spin (2A2S) Hamiltonian in place of the maximally entangled state and examine its performance.  We find that the version of the protocol with 2A2S squeezing well-approximates the maximally entangled state, such that spin averages can be remotely prepared.  We evaluate the errors that are introduced by using 2A2S squeezed states, and find that it decreases with the ensemble size.  With post-selection, errors can be systematically decreased further. 
\end{abstract}

\maketitle

\section{\label{sec1}Introduction}

Entanglement is one of the fundamental distinguishing characteristics of quantum mechanics with respect to classical physics \cite{einstein1935can,horodecki2009quantum}, and is considered a resource in the modern context of quantum information science \cite{chitambar2019quantum}.  It plays a fundamental role in non-trivial quantum protocols such as quantum teleportation \cite{bennett1993teleporting}, and its generation is considered to be one of the essential capabilities when constructing a quantum computer \cite{ladd2010quantum}.  There are numerous different physical systems where entangled states have been prepared experimentally, such as superconductors \cite{berkley2003entangled,dicarlo2010preparation}, photons \cite{yin2017satellite,ren2017ground}, quantum dots \cite{akopian2006entangled,bayer2001coupling,ishida2013photoluminescence}, NV centers \cite{neumann2008multipartite}, neutral atoms  \cite{isenhower2010demonstration,picken2018entanglement}, trapped ions \cite{schindler2013quantum}. While 
entanglement is most often associated with the microscopic world, it has been also shown to be present in quantum many-body systems \cite{amico2008entanglement,its2005entanglement,zhang2005thermal,radhakrishnan2017quantum,pezze2019heralded}. For macroscopic systems, particularly atomic ensembles, most of the work to date has been focused on single atomic ensembles, where the entanglement exists between atoms in the {\it same} ensemble \cite{gross2012spin}.  

 Entanglement is fundamental to quantum squeezing, which allows a way to overcome the standard quantum limit \cite{kitagawa1993squeezed,wineland1994squeezed,sorensen2001many,pezze2018quantum}. Squeezing has many potential applications in quantum metrology that has triggered many experiments so far \cite{hald1999spin,kuzmich2000generation,esteve2008squeezing,riedel2010atom,zhang2014quantum,krauter2011entanglement}. 
Two well-known types of squeezed states on single ensembles are the one-axis and two-axis countertwisting squeezed states \cite{kitagawa1993squeezed}.  The first demonstration of entanglement {\it between} two atomic ensembles was performed by Julsgaard, Kozhekin, and Polzik \cite{julsgaard2001experimental}, where a two-mode squeezed state was produced, mediated by an optical pulse. Continuous variables teleportation \cite{krauter2013deterministic} and spin-wave teleportation \cite{bao2012quantum} was accomplished based on the generation of entanglement between spatially separated atomic ensembles.  For ultracold atoms, experiments towards generating entanglement between two atomic clouds is currently being pursued.  Entanglement between two spatial regions of the same Bose-Einstein condensate (BEC) was reported simultaneously by three groups \cite{kunkel2018spatially,lange2018entanglement,fadel2018spatial}. 
The generation of entanglement between two BECs has been investigated theoretically in numerous works \cite{PhysRevA.74.022312,pyrkov2013,abdelrahman2014coherent,rosseau2014,hussain2014,oudot2017optimal,Jing_2019}, mainly focusing on the generation of the two-spin version of the one-axis countertwisting Hamiltonian, among others \cite{olov,idlas2016,ortiz2018adiabatic}. Such an interaction can be considered the two-ensemble version of the one-axis squeezed state, due to the similar form of the generating Hamiltonian.  We call this state the one-axis two-spin (1A2S) squeezed state, which have been studied in detail in works such as Refs. \cite{byrnes2013fractality,kurkjian2013spin}, and have been shown to exhibit interesting properties such as fractal pattern of entanglement. 

Recently, the two-ensemble version of two-axis countertwisting squeezed state was investigated \cite{kitzinger2020}, which we call the two-axis two-spin (2A2S) squeezed state.   In contrast to the 1A2S squeezed state which produces correlations in two pairs of spin variables, the 2A2S squeezed states produces correlations between all three spin variables \footnote{For 1A2S states in the short time limit, the dominant correlations are of the form $ S^z_A - S^y_B $ and $ S^y_A - S^z_B $ \cite{byrnes2013fractality}.  Meanwhile, the 2A2S states have correlations of the form $ S^x_A + S^x_B $, $ S^y_A - S^y_B $,$ S^z_A - S^z_B $ after a suitable basis rotation \cite{kitzinger2020}.}. The lack of full correlations of the 1A2S state is evident in past approaches for using the state as a basis for quantum teleportation 
\cite{pyrkov2014quantum,pyrkov2014full}.  In Ref. \cite{pyrkov2014quantum}, only spin coherent states around the equator of the Bloch sphere could be successfully teleported.  Ref. \cite{pyrkov2014full} overcame this limitation by using two auxiliary ensembles to produce correlations in more spin directions, to achieve teleportation for any state on the Bloch sphere.  This makes the 2A2S squeezed states potentially a better candidate from a quantum information perspective \cite{byrnes2012macroscopic,byrnes2015macroscopic,semenenko2016implementing,mohseni2019error}, which have mainly considered 1A2S squeezed states in the past.    

In this paper, we propose and analyze a scheme for performing remote state preparation (RSP) between two BECs based on the 2A2S squeezed state. RSP  \cite{lo2000classical,pati2000minimum,bennett2001remote,berry2003optimal,leung2003oblivious,peng2003experimental,ye2004faithful,peters2005remote,dakic2012quantum} is a protocol that is similar in many ways to quantum teleportation, but possesses several key differences. The aim of the original version of the protocol is to prepare a desired state of a qubit using the help of entanglement and classical communication. The aim in our case is to prepare desired spin averages remotely on a BEC via the use of shared entanglement. Such a protocol is interesting in context of recent experiments that are being developed currently relating to bimodal BEC-BEC entanglement.  While protocols for teleportation using BECs have been proposed, these are still out of reach of current experiments since they involve entanglement between three \cite{pyrkov2014quantum} and four \cite{pyrkov2014full} bimodal BEC clouds. In RSP, only two atomic clouds are required, hence is a simpler goal in the near-term.  In the standard qubit version of the RSP protocol, only equatorial states on the Bloch sphere are usually considered, since there is no classical correction that can correct a more general state.  We shall consider a slightly more general version of RSP where arbitrary states on the Bloch sphere are prepared on the target side.  This will highlight the full set of spin correlations that are present in 2A2S squeezed state.  This allows for the transfer of full coordinates on the Bloch sphere, up to a reflection on the $ z $-axis.  The ambiguity regarding the reflection occurs due to the limitation of the original qubit RSP protocol, as there is no anti-unitary classical correction \cite{bennett2001remote}.   

We note that since a two-component BEC is a finite-dimensional system, in principle it is possible to apply existing RSP protocols for qudits \cite{bennett2001remote,zeng2002remote}. In this sense, our aim of remotely preparing spin averages is a more restrictive goal than preparing a general state. The main problem here is that it is assumed that a measurement in an arbitrary basis is available. In the case of BECs, only a limited set of measurements are realistic, typically Fock state measurements.  Additionally, unitary operations corresponding to Hamiltonian with low powers of total spin operators are usually only available. When designing our protocol we constrain ourselves such that: (i) only operations involving low powers of total spin operators are used;  (ii) only measurements in Fock basis are used; (iii) decoherence-sensitive encodings are not used (see Ref. \cite{timquantumoptics2020} for more details about the approach). We choose this because of the limited measurements and operations that are available to a BEC. As spin coherent states are robust in the presence of decoherence, they fulfil the requirement (iii). Furthermore, this can be considered a protocol that fits within the scheme of spinor quantum computing \cite{timquantumoptics2020}, where qubit information is encoded redundantly.  This has been shown to be an  error suppressed encoding of the original qubit encoding \cite{mohseni2019error}. These design principles allow for a robust protocol that should be implementable in future experiments.

We finally comment on similarities and differences to existing works.  
In Ref. \cite{ye2004faithful}, the RSP protocol was generalized to the high dimensional case with non-maximally entangled states. The 2A2S squeezed state produces non-maximally entangled states hence a question is where the approach of Ref. \cite{ye2004faithful} could be used.  In our case, the entangled resource that will be used is the optimally squeezed 2A2S state, which is not a maximally entangled state. The successful preparation of the given state in Ref. \cite{ye2004faithful} depends upon the matching state structure of the entangled and desired state.  In our case, since this structure is not there, it does not appear possible to apply this approach in a way that practically feasible in the sense described above. Finally, another potential option is to work under the Holstein-Primakoff approximation \cite{krauter2013deterministic} and use the approach of Ref. \cite{kurucz2005continuous}.  This is however also not applicable in our case, since in a similar way to Refs.  \cite{pyrkov2014quantum,pyrkov2014full}, our aim is to be able to prepare a state at any position on the Bloch sphere, not only those which are in the vicinity of the polarized spin, as demanded by the Holstein-Primakoff approximation \cite{julsgaard2001experimental}.

The paper is structured as follows: in Sec. \ref{sec2}, we present the RSP protocol for the ideal cases of a single qubit and BECs prepared with maximally entangled Einstein-Podolsky-Rosen (EPR) states.  We then explain our proposed RSP protocol for 2A2S squeezed states in Sec. \ref{sec3}. In Sec. \ref{sec4}, we analyze the performance of the RSP protocol with 2A2S squeezed states.  The conclusions are given in Sec. \ref{sec5}.

%

\section{Remote state preparation: ideal cases}
\label{sec2}

We first examine two ideal RSP protocols that will be the foundation for our full RSP protocol to be discussed in later sections.  The first examines the extended RSP protocol for preparing a state of an qubit with arbitrary coordinates on the Bloch sphere.  The second introduces the ideal version of the RSP protocol for spin ensembles, where the equivalent operation to the qubit RSP is performed.

\subsection{Qubit remote state preparation}

 Let us recall the RSP protocol for the qubit case \cite{lo2000classical,pati2000minimum,bennett2001remote}.  The continuous variable version of RSP is given in \cite{kurucz2005continuous}.  
In the first step, one prepares a maximally entangled state, for example the state $ ( | 0 \rangle_A | 0 \rangle_B - | 1 \rangle_A | 1 \rangle_B)/\sqrt{2} $.  Then Alice (in possession of the first qubit), performs a measurement in the basis
 \begin{align}
| A_0 \rangle & = e^{-i \sigma^z( \pi - \phi)/2} e^{-i \sigma^y \theta /2} | 0 \rangle \propto  \cos \frac{\theta}{2} | 0 \rangle - e^{-i \phi} \sin  \frac{\theta}{2}   | 1 \rangle  \nonumber \\
| A_1 \rangle & = e^{-i \sigma^z( \pi - \phi)/2} e^{-i \sigma^y \theta /2} | 1 \rangle \propto  \sin  \frac{\theta}{2}  | 0 \rangle + e^{-i \phi} \cos  \frac{\theta}{2} | 1 \rangle  ,
\label{qubitbasismeasure}
\end{align}
where $ \sigma^x,\sigma^y,\sigma^z $ are the $ 2 \times 2 $ Pauli matrices.  Alice then measures in the $ |A_0 \rangle, |A_1 \rangle $ basis and informs Bob (in possession of the second qubit) of the result. 

For the case that Alice obtains $ |A_0 \rangle $, which occurs with probability $ p =1/2$, Bob obtains the state
\begin{align}
  | B_0 \rangle =  \cos  \frac{\theta}{2} | 0 \rangle +  e^{i \phi} \sin  \frac{\theta}{2}  | 1 \rangle . 
    \label{qubitrspcase0}
\end{align}
This is a completely general state on the Bloch sphere, which completes the RSP for this case.  For the case that Alice obtains $ |A_1 \rangle $, which occurs with probability $ p =1/2$, Bob obtains the state $ \sin  \frac{\theta}{2} | 0 \rangle  -  e^{i \phi} \cos  \frac{\theta}{2}  | 1 \rangle $. The phase between the states can be fixed to be the same as (\ref{qubitrspcase0}) by Bob conditionally applying the unitary $ e^{-i\sigma^z \pi/2} $, to give the state
\begin{align}
  | B_1 \rangle =   \sin  \frac{\theta}{2} | 0 \rangle  +  e^{i \phi} \cos  \frac{\theta}{2}  | 1 \rangle .
    \label{qubitrspcase1}
\end{align}
There is however no unitary operation that can turn the state (\ref{qubitrspcase1}) into (\ref{qubitrspcase0}) for general $ \phi $ \cite{bennett2001remote}.  For this reason, only equatorial states $ \theta = \pi/2 $ are usually considered in the RSP protocol, where  (\ref{qubitrspcase1}) is the same state as (\ref{qubitrspcase0}) \cite{pati2000minimum,berry2003optimal,berry2004resources}.  

Another way to view this is in terms of expectation values of spin operators.  For the case that Alice obtains the state $ |A_0 \rangle $, the expectation values of the Pauli operators are 
\begin{align}
    \langle B_0 | \sigma^x | B_0 \rangle & = \sin \theta \cos \phi \nonumber \\
    \langle B_0 | \sigma^y | B_0 \rangle & = \sin \theta \sin \phi \nonumber \\
    \langle B_0 | \sigma^z | B_0 \rangle & = \cos \theta ,
    \label{rspbob0}
\end{align}
which are the expectation values in terms of the standard Bloch sphere angles.  For the state (\ref{qubitrspcase1}), we however have
\begin{align}
    \langle B_1 | \sigma^x | B_1 \rangle &= \sin \theta \cos \phi  \nonumber \\
    \langle B_1 | \sigma^y | B_1 \rangle &= \sin \theta \sin \phi \nonumber \\
    \langle B_1 | \sigma^z | B_1 \rangle &= -\cos \theta .
        \label{rspbob1}
\end{align}
Hence Bob receives the target state up to a reflection Bloch sphere about the $ x$-$y $ plane.  Since both Alice and Bob are aware of the measurement outcome, one way to deal with the additional minus sign is to add the extra sign during classical post-processing. In this case, Bob takes note that his state is a reflected version of the intended state if Alice informs that she obtained $ |A_1\rangle $, and accounts for this in his  subsequent operations.  

Following the procedure above, Alice can remotely prepare Bob's qubit in a desired quantum state. While the protocol has similarities with teleportation, there are several differences. Firstly, the state to be prepared on Bob's side is known to Alice in advance, unlike teleportation where it is in principle unknown to Alice and Bob. Secondly, it only involves two qubits, rather than three, such that no Bell measurement is required.  Finally, only one bit of classical information is sent from Alice to Bob, in contrast to teleportation where two bits are required.

\subsection{Remote state preparation using a spin-EPR state}
\label{sec:rspspinepr}

We now introduce a variant of the RSP protocol suitable for spin ensembles and BECs that allows Alice to prepare an arbitrary state on Bob's side with the same Bloch sphere parameters as that seen in (\ref{rspbob0}) and (\ref{rspbob1}). 

Alice and Bob are each in possession of a two-component BEC or atomic ensemble with $ N $ atoms respectively. Working in the symmetric subspace of the atomic ensembles, our formalism equally applies to either thermal atomic ensembles or BECs (see Sec. 5.10 of Ref.  \cite{timquantumoptics2020} for a full discussion). The exact mathematical equivalence of the internal spin states of BECs and atomic ensembles in the symmetric subspace allows us to handle both cases simultaneously. For brevity, we will call the atomic ensembles held by Alice and Bob as ``BECs'', although it should be understood that our formalism equally holds for thermal ensembles. The bosonic annihilation operators for the two component BEC are denoted as $ a, b $. Any state of the BEC can then be expanded in terms of the Fock basis
\begin{align}
    | k \rangle = \frac{1}{\sqrt{k!(N-k)!}} (b^\dagger)^k (a^\dagger)^{N-k} | \text{vac} \rangle 
    \label{fockbasis}
\end{align}
for each BEC, where $ |\text{vac} \rangle $ is the vacuum state with no particles. States involving superposition of the states $ a,b$ can also form Fock states, which we define
 \begin{align}
| k \rangle^{(\theta, \phi)} = U^{(\theta, \phi)} | k \rangle ,  
\label{thetaphifock}
\end{align}
where 
\begin{align}
    U^{(\theta, \phi)} = e^{-i S^z \phi/2} e^{-i S^y \theta/2} 
\end{align}
is the unitary operation that rotates a state from the north pole of the Bloch sphere to spherical coordinates $ (\theta, \phi)$. Here the spin operators are Schwinger boson (total spin) operators for Alice and Bob respectively, defined as
\begin{align}
    S^x & =b^\dagger a+a^\dagger b \nonumber \\
    S^y & =-ib^\dagger a+ia^\dagger b \nonumber \\
     S^z & =b^\dagger b-a^\dagger a
\end{align}
The commutation relation for spin operators is given by 
\begin{align}
    [S^j,S^k]=2i\epsilon_{jkl}S^l
\end{align}
where $\epsilon_{jkl}$ is the Levi-Civita symbol and $ j,k,l \in \{ x,y,z \} $.
Explicit expressions for the matrix elements of the rotated Fock states are given in Appendix \ref{app:prob}.  

The first step of the protocol involves preparing an entangled state.  Let us consider the maximally entangled state 
\begin{align}
| \text{EPR}_- \rangle = \frac{1}{\sqrt{N+1}} \sum_{k=0}^N (-1)^k |k\rangle_A  |k\rangle_B .
\label{eprzbasis}
\end{align}
This state was considered in Ref. \cite{kitzinger2020} and was shown to have similarities to the 2A2S squeezed state. Analogously to (\ref{qubitbasismeasure}), the next step is then to measure in the basis
 \begin{align}
| k \rangle^{(\theta, \pi - \phi)} =U^{(\theta, \pi - \phi)}  | k \rangle .  
\end{align}
This can be done by Alice applying the unitary rotation 
\begin{align}
    {U^{(\theta, \pi - \phi)}}^\dagger   = e^{i S^y_A \theta/2}  e^{i S^z_A (\pi - \phi) /2} 
    \label{aliceunitary}
\end{align}
then performing a measurement in the Fock basis (\ref{fockbasis}). Such Fock state measurements are readily achievable using the current state of the art technology for BECs \cite{riedel2010atom,bohi2009coherent,byrnes2015macroscopic}.  

One of the important features of the state (\ref{eprzbasis}) is that it can be algebraically manipulated into the form \cite{kitzinger2020}
\begin{align}
 | \text{EPR}_{-} \rangle & = \frac{1}{\sqrt{N+1}} \sum_{k=0}^N | k \rangle_A^{(\theta,\pi - \phi)} | k \rangle_B^{(\theta,\phi)}  	 . \label{spinEPRthetaphi}
 \end{align}
 Using this result, applying the unitary rotation (\ref{aliceunitary}), we have
 \begin{align}
   {U^{(\theta, \pi - \phi)}}^\dagger   | \text{EPR}_{-} \rangle  
    & = \frac{1}{\sqrt{N+1}} \sum_{k=0}^N | k \rangle_A | k \rangle_B^{(\theta,\phi)}.  
    \label{intermedstate}
    \end{align}
Alice's projection on a particular Fock state $| k \rangle_A $, which occurs with probability $ p_k = 1/(N+1)$, gives the state $ | k \rangle_B^{(\theta,\phi)} $ on Bob side. Projections on Fock states can be performed experimentally using various techniques such as spin-selective absorption imaging \cite{pezze2018quantum}. 

The final step is to apply a $ \pi $-rotation around the $ z $-axis
\begin{align}
U_k^C = \left\{
\begin{array}{cc}
e^{-iS^z_B \pi/2}  & \hspace{1cm} k < N/2 \\
I  & \hspace{1cm} k\ge N/2 .
\end{array}
\right.  \label{classicalcorr}
\end{align}
This is the analogous step to that performed to obtain the state (\ref{qubitrspcase1}) in the qubit case.  The final state held by Bob at the end of the RSP protocol is then
\begin{align}
    | \Psi_k^{\text{ideal}} \rangle 
 = \left\{
\begin{array}{cc}
   | k \rangle_B^{(\theta,\phi+ \pi )}  & \hspace{1cm}  k< N/2  \\
   | k \rangle_B^{(\theta,\phi )}    & \hspace{1cm}  k\ge  N/2 .
\end{array}
\right.
    \label{finalstatespinepr}
\end{align}
The above procedure completes the RSP as desired. 

We can see that (\ref{finalstatespinepr}) achieves the desired aim by evaluating the expectation values of the spin operators.  For the case $ k \ge N/2 $, we have
\begin{align}
\langle  \Psi_k^{\text{ideal}}& 
  | S^x| \Psi_k^{\text{ideal}} \rangle 
 =
\langle k | {U^{(\theta,\phi)}}^\dagger S^x U^{(\theta,\phi)} | k \rangle \nonumber \\
&=\langle k  | ( \cos \theta \cos \phi S^x  - \sin \phi S^y + \sin \theta \cos \phi S^z ) | k \rangle  \nonumber \\
& = (2k-N) \sin \theta \cos \phi  .  
\label{sxavkplus}
\end{align}
where the transformation of the spin operators was used, and the $ S^x, S^y $ terms do not contribute since they are off-diagonal.  Similarly, we evaluate
\begin{align}
\langle \Psi_k^{\text{ideal}}& | S^y| \Psi_k^{\text{ideal}} \rangle 
 =
\langle k | {U^{(\theta,\phi)}}^\dagger S^y U^{(\theta,\phi)} | k \rangle \nonumber \\
& = \langle k  | ( \cos \theta  \sin \phi  S^x +\cos \phi S^y + \sin \theta \sin \phi  S^z ) | k \rangle  \nonumber \\
& = (2k-N) \sin \theta \sin \phi ,  
\label{syavkplus}
\end{align}
and 
\begin{align}
\langle \Psi_k^{\text{ideal}}  | S^z| \Psi_k^{\text{ideal}} \rangle 
&=
\langle k | {U^{(\theta,\phi)}}^\dagger S^z U^{(\theta,\phi)} | k \rangle \nonumber \\
& = \langle k  | ( \cos \theta S^z - \sin \theta  S^x ) | k \rangle  \nonumber \\
& = (2k-N) \cos \theta.  
\label{szavkplus}
\end{align}
We can see that (\ref{sxavkplus}),  (\ref{syavkplus}), (\ref{szavkplus}) are the analogous result to (\ref{rspbob0}) for qubits.  There is a common factor of $ 2k-N > 0 $ which can be eliminated by normalizing the Bloch vector with the factor
\begin{align}
    {\cal N} \equiv
    \sqrt{ \langle S^x \rangle^2  + \langle S^y \rangle^2 + \langle S^z \rangle^2} 
    = | 2k -N | .  \label{normfactor}
\end{align}

For the $ k < N/2$ case, the expectation values can be similarly evaluated as
\begin{align}
\langle \Psi_k^{\text{ideal}}  | S^x| \Psi_k^{\text{ideal}} \rangle 
 &=
(N- 2k) \sin \theta \cos \phi  \nonumber \\  
\langle \Psi_k^{\text{ideal}}  | S^y| \Psi_k^{\text{ideal}} \rangle  &=
(N- 2k) \sin \theta \sin \phi  \nonumber \\  
\langle \Psi_k^{\text{ideal}}  | S^z| \Psi_k^{\text{ideal}} \rangle  &=
 - (N- 2k) \cos \theta 
\label{sxsyszavkminus}
\end{align}
which is the analogous result to (\ref{rspbob1}). The factors of $ N - 2k > 0 $ can be eliminated by again dividing by (\ref{normfactor}). 
The above result can be summarized for all $k $ as
\begin{align}
\langle \Psi_k^{\text{ideal}} | S^j |  \Psi_k^{\text{ideal}} \rangle = \left\{
\begin{array}{cc}
|2k-N| \sin \theta \cos \phi   & j=x \\
|2k-N|  \sin \theta \sin \phi   & j=y \\
(2k-N) \cos \theta    & j=z \\
\end{array}
\right.
\label{idealspinexp}
\end{align}
 We again see that $ \langle S^z \rangle $ has an extra minus sign for $ k < N/2 $ which cannot be eliminated using a unitary rotation.  

The above protocol completes the aim of the RSP, where the Bloch sphere parameters $ (\theta, \phi)$ are prepared on Bob's side by Alice. As with the standard qubit case, only one bit of classical information is required to perform the classical correction step. For the projection outcome $ k < N/2 $, there is an extra minus sign which cannot be eliminated, similar to the qubit case.  This can be handled either by Bob performing classical post-processing on his results, or one can restrict Alice's measurements to $ \theta = \pi/2 $ which removes the issue by setting $ \langle S^z \rangle = 0 $. We note that the protocol succeeds except for the outcome $ k = N/2$, where all expectation values are zero.  The failure probability is $ 1/(N+1)$ for even $ N $ and zero for odd $ N $.  For large $ N $ this is a rare outcome, and can be considered an isolated case.  In this way, all measurement outcomes except for $ k = N/2 $ in Fig. \ref{fig2}(a) can be utilized as successful outcomes in the RSP protocol.  For other qudit protocols the {\it success} probability is $ 1/(N+1)$ \cite{bennett2001remote}, hence this is a considerable improvement in comparison.  This is due to the more restricted aim of preparing the Bloch sphere parameters $ (\theta, \phi)$, whereas in Ref. \cite{bennett2001remote} is to prepare a general quantum state.  The aim of preparing the Bloch angles is consistent with the approach of Ref. \cite{mohseni2019error} where the BEC acts as an error suppressed encoding of qubit states.

\section{Remote state preparation protocol with the two-axis two-spin squeezed state}
\label{sec3}

In Sec. \ref{sec:rspspinepr} we introduced a RSP protocol based on spin-EPR states.  While this is satisfactory as a protocol in terms transferring a state with Bloch sphere angles $ (\theta, \phi)$, preparation of the spin-EPR state is non-trivial using current experimental techniques.  In Ref. \cite{kitzinger2020}, it was found that the 2A2S squeezed state can approximate the spin-EPR state at particular evolution times. Our strategy will thus be to replace the spin-EPR state with the 2A2S squeezed state and proceed with the RSP protocol as described in the previous section.  In this section, we describe the RSP protocol using 2A2S squeezed states.

\subsection{The two-axis two-spin (2A2S) squeezed state}
In this section we briefly review the two-axis two-spin squeezed state that will be used as the entangled state for the remote state preparation.  This is discussed in more detail in Ref. \cite{kitzinger2020}. The 2A2S Hamiltonian is defined as 
\begin{equation}
    H= \frac{J}{2} ( S_{A}^{x}S_{B}^{x}-S_{A}^{y}S_{B}^{y} ) =J (S_{A}^{+}S_{B}^{+}+S_{A}^{-}S_{B}^{-}) ,
    \label{twoaxiscosqueezedham}
\end{equation}
where the raising and lowering operators are defined by, 
\begin{align}
    S^+_{j} & = (S^x_{j} + i S^y_{j})/2 =    b^\dagger_{j} a_{j} \nonumber \\
    S^-_{j} & = (S^x_{j} - i S^y_{j})/2 =   a^\dagger_{j} b_{j},   
\end{align}%
where $ j \in \{A,B\}$ and $ J $ is an energy constant.  The Hamiltonian (\ref{twoaxiscosqueezedham}) is the natural generalization of the two-axis squeezing Hamiltonian \cite{kitagawa1993squeezed,gross2012spin} acting on two spin ensembles.  For the single ensemble case, the two-axis squeezing Hamiltonian reduces the quantum noise to a greater extent than one-axis squeezing. 

The 2A2S Hamiltonian is applied to two atomic ensembles initially prepared in the maximally $+S^z$-polarized state to create the entangled state.  The state that we will consider is
\begin{align}
    | \psi (\tau )  \rangle = e^{-i (S_{A}^{+}S_{B}^{+}+S_{A}^{-}S_{B}^{-} ) \tau} | N  \rangle_A | N \rangle_B ,
    \label{twoaxiscosqueezed}
\end{align}
where the initial states are Fock states that satisfy $S^z | N  \rangle = N | N \rangle $ according to (\ref{fockbasis}), and we defined the dimensionless time parameter $\tau = J t/\hbar $.  

The 2A2S squeezed state (\ref{twoaxiscosqueezed}) has EPR-like correlations, in a similar way to two-mode squeezed states in continuous variables optics \cite{braunstein2005quantum}. According to our Hamiltonian phase convention, the relevant squeezed variables are the rotated operators $(S^y \pm S^x)/\sqrt{2} $ \cite{kitzinger2020}.  In order to account for this rotation, it is convenient in our case to apply a phase rotation to the state (\ref{twoaxiscosqueezed}), such that we work instead with the state
\begin{align}
    | & \Psi(\tau) \rangle  = e^{i S^z_A \pi/8} e^{i S^z_B \pi/8}  | \psi (\tau )  \rangle \nonumber \\
     & = e^{i S^z_A \pi/8} e^{i S^z_B \pi/8}  e^{-i (S_{A}^{+}S_{B}^{+}+S_{A}^{-}S_{B}^{-} ) \tau} | N  \rangle_A | N \rangle_B .  \label{2a2sstate}
\end{align}
Using a Holstein-Primakoff approximation, and for short times $ \tau < \ln (4N)/2N
 $ \cite{kitzinger2020}, the variance of the state (\ref{2a2sstate}) follows the relations 
\begin{align}
    \text{Var} (S^{x}_{A}+ S^{x}_{B}) & = 2N e^{-2N \tau} \nonumber \\
       \text{Var} (S^{y}_{A}- S^{y}_{B}) & = 2N e^{-2N \tau} \nonumber \\
  \text{Var} (S^{z}_{A}- S^{z}_{B}) & = 0 .  
\end{align}
Under the full evolution (\ref{twoaxiscosqueezed}), the first two variances reach a minimum at the optimal squeezing time and then increases again.  In the vicinity of this time, it was observed in Ref. \cite{kitzinger2020} that the state (\ref{2a2sstate}) has a high fidelity with the state (\ref{eprzbasis}), such that we may approximate
\begin{align}
    |\text{EPR}_- \rangle \approx | & \Psi(\tau_{\text{opt}}) \rangle.  
\end{align}
In this paper, we define the evolution time $ \tau_{\text{opt}} $ such that the maximum fidelity with the spin-EPR state is achieved.  The optimum times for each $ N $ are provided in Ref. \cite{kitzinger2020}. We use the above state in place of the spin-EPR state to accomplish RSP.

\subsection{Remote state preparation protocol}
\label{sec:protocol}

Here we summarize, for the sake of clarity, the RSP protocol using 2A2S squeezed states, as developed in the previous sections.  The protocol follows the sequence:
\begin{enumerate}
\item Prepare two BECs in the maximally $ + S^z $ polarized state, and apply the 2A2S Hamiltonian (\ref{twoaxiscosqueezedham}) for a time $ \tau_{\text{opt}} $, according to (\ref{twoaxiscosqueezed}). The optimal time $ \tau_{\text{opt}} $ is the time that optimizes the fidelity with the spin-EPR state, given in Ref. \cite{kitzinger2020}.  
\item Apply the transformations $ e^{i S^z_A \pi/8} e^{i S^z_B \pi/8} $ such as to produce correlations that are similar to the spin-EPR state (\ref{eprzbasis}).  
\item Apply a unitary operation (\ref{aliceunitary}), i.e. $  {U^{(\theta, \pi - \phi)}}^\dagger = e^{i S^y_A \theta/2} e^{i S^z_A (\pi-\phi) /2} $ on Alice's BEC.   
\item Alice measures in the $S^z $-basis Fock states $|k \rangle $ and tells Bob the binary result of whether $ k < N/2$ or $ k \ge N/2 $.  
\item Bob applies the unitary (\ref{classicalcorr}), i.e. if $ k < N/2 $ then applies $e^{-i S^z \pi/2} $, and otherwise does nothing.   
\end{enumerate}

The above produces an approximation to expectation values (\ref{idealspinexp}).   

\section{Numerical analysis of the remote state preparation protocol}
\label{sec4}

In this section, we analyze the protocol that is provided in Sec. \ref{sec:protocol}.  In general, the state (\ref{twoaxiscosqueezed}) cannot be written analytically and hence one must evolve the state numerically to obtain the wavefunction. We examine quantities such as the probability distribution, spin averages, Wigner functions, and the error of the protocol due to the usage of the 2A2S squeezed states.

\subsection{Probability distribution}

We first examine the probability distribution for RSP state of the protocol.  This is defined by
\begin{align}
    P_k = \langle \Psi_k | \Psi_k \rangle \label{probability}
\end{align}
where
\begin{align}
  |\Psi_k\rangle = U_k^C |k\rangle \langle k|_A {U^{(\theta, \pi-\phi)}}^\dagger | \Psi(\tau_{\text{opt}}) \rangle 
 \label{rspstate}
\end{align}
is the resulting unnormalized state after step 5 in the RSP protocol on Bob's side. The expression for the rotation of the Fock states are given in the Appendix \ref{app:prob}.
The measurement $ | k \rangle \langle k |_A $ is performed on Alice's subspace and $U_k^C $ is performed on Bob's subspace.    

The probability distribution is independent of $ \phi $ and only depends upon $ \theta $. To see this, let us write (\ref{twoaxiscosqueezed}) as
\begin{align}
    | \psi (\tau) \rangle = \sum_k \psi_k | k \rangle_A | k \rangle_B,
\end{align}
where we used the fact that only equal Fock number states are generated by the 2A2S Hamiltonian. The RSP state (\ref{rspstate}) then can be written 
\begin{align}
  |  \Psi_k \rangle = |k\rangle_A \otimes \sum_{k'} & \psi_{k'} e^{i (2k'-N)(\frac{3 \pi-\phi}{4}-\frac{\Theta(N-2k)\pi}{2})} \nonumber\\
  & \times \langle k|_A e^{i S^y_A \theta/2} |k'\rangle_{A} |k'\rangle_{B}  .
\end{align}
where $ \Theta(N-2k) $ is the Heaviside step function.
The probability distribution is then given as
\begin{align}
    P_k(\theta) = \sum_{k'}|\psi_{k'}|^2 |\langle k| e^{i S^y \theta/2} |k'\rangle|^2.
\end{align}
which makes it clear that the $ \phi $-dependent phase terms cancel out.

\begin{figure}[t]%
    \includegraphics[width=\linewidth]{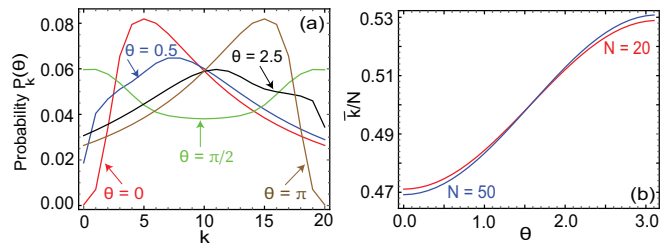}
    \caption{(a) RSP probability distributions of measurements on Alice’s qubits $P_k(\theta) $ for $ N = 20 $, time $ \tau_{\text{opt}} = 0.1214 $, (b) Average of the normalized probability distribution for $ N = 20 $, time $ \tau_{\text{opt}} = 0.1214 $ and $ N = 50 $, time $ \tau_{\text{opt}} = 0.0586 $. }%
    \label{fig2}%
\end{figure}

In our calculations, we generally focus upon the case with $ N \gg 1$, appropriate for BECs which may typically contain atoms beyond $ N = 10^3$ \cite{bohi2009coherent}.  We typically take $ N $ large enough such that the effects of $ N $ are not significant. In Fig. \ref{fig2} we show the probability distribution for various values of $ \theta $. It is clear that under a transformation of $ \theta \to \pi - \theta $, the probability distribution transforms as $ k \to N-k $. The extremal values of angles such as $ \theta = \pi $ have a low probability outcome for $ k = N $, and similarly there is a low probability outcome for $ k = 0 $ around $ \theta = 0 $. For larger $ N $, the probability distribution has a similar characteristics.

We also plot the average of the normalized probability distribution defined by
\begin{align}
    \bar{k} = \sum_k k P_k(\theta). 
    \label{Avg}
\end{align}
The numerical results are obtained for $ \bar{k}/N $ for two $ N $ values as shown in Fig. \ref{fig2}(b). We see that the average $ k $ outcome is very close to $ N/2 $ for a wide range of $ \theta $, with a weak cosine dependence.  Thus although the distribution is biased by the target $ \theta $ state, the measurement outcome generally has a broad range of outcomes in $ k $.

\subsection{Spin Averages after Alice's measurement}
\label{sec:spinavs}

\begin{figure*}[t]%
    \includegraphics[width=\textwidth]{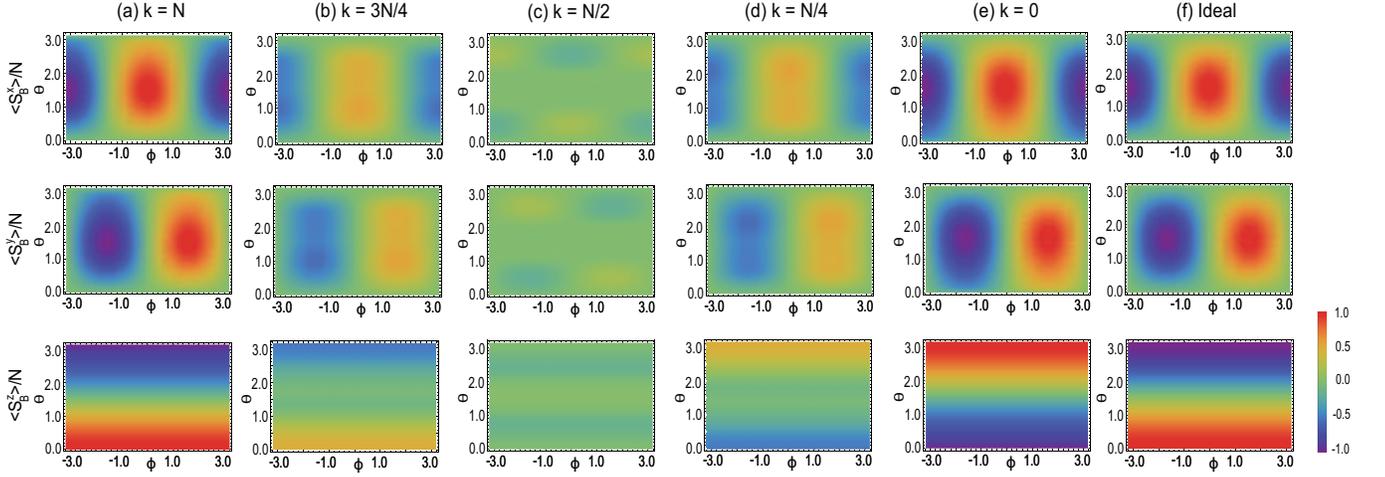}%
    \caption{Spin averages $ \langle S^x_B \rangle, \langle S^y_B\rangle, \langle S^z_B \rangle $  for Bob's BEC according to (\ref{spinaverages}) for different projective measurements (a) $ k = N $, (b) $ k = 3N/4 $, (c) $ k = N/2 $, (d) $ k = N/4 $, (e) $ k = 0 $  for $ N = 20 $,  $ \tau_{\text{opt}} = 0.1214 $. (f) Ideal case (\ref{idealspinexp}) with $ k = N $. }%
    \label{fig3}%
\end{figure*}

We measure the spin averages of the RSP state (\ref{rspstate}) defined by
\begin{align}
    \langle S^j_B \rangle = \frac{\langle\Psi_k|S^j_B|\Psi_k\rangle}{\langle \Psi_k | \Psi_k \rangle}  \label{spinaverages}
\end{align} 
where $ j \in \{ x, y, z \} $. Figure \ref{fig3} shows the spin average density plot over the entire Bloch sphere. The expectation values for an ideal remote state preparation are given in (\ref{idealspinexp}) and are plotted in Fig. \ref{fig3}(f) for comparison.   For the projection outcome $ k = 0 $, we see that the spins distribution is completely flipped for the $ S^z $ variable in comparison with Fig. \ref{fig3}(f) and is in agreement with (\ref{sxsyszavkminus}). Moreover, the spin amplitudes are diminished for outcomes near $ k = N/2 $, where spin averages are close to zero as in Fig. \ref{fig3}(c). 

\begin{figure}[t]%
    \includegraphics[width=\linewidth]{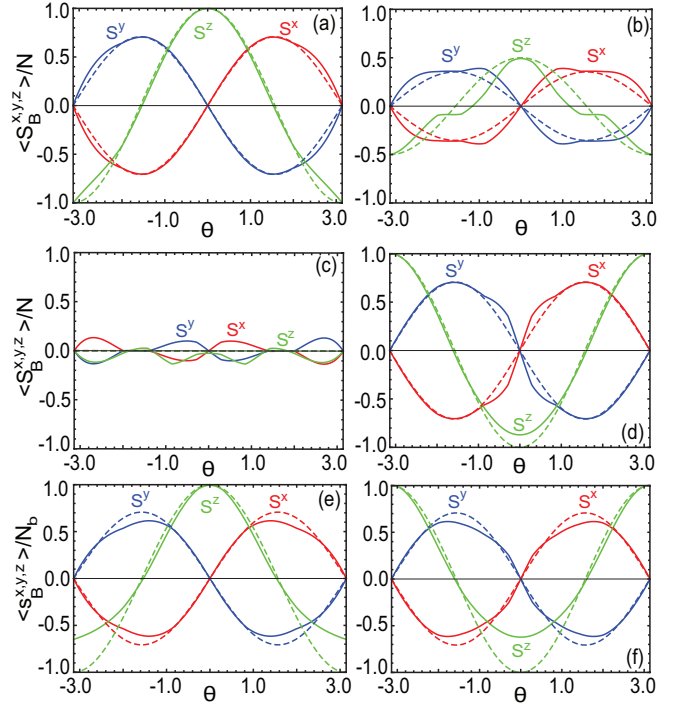}
    \caption{Spin averages (\ref{spinaverages}) for Bob's state corresponding to Alice's outcomes in RSP protocol for $N = 20 $, time $ \tau_{\text{opt}} = 0.1214 $, $ \phi = -\pi/4 $ (a) $ k = N $, (b) $ k = 3N/4 $, (c) $ k = N/2 $, (d) $ k = 0 $. Spin averages (\ref{stat.}) for BECs prepared in a statistical mixture of different atom numbers with (e) $ k = N_A $, (f) $ k = 0 $ about mean atom number $ \bar{N} = 20 $ for Gaussian width $ \sigma_0 = 2\sqrt{ \bar{N}} $. 
    Dotted lines are spin averages for the ideal case (\ref{idealspinexp}). }%
    \label{fig4}%
\end{figure}

Taking a closer look at the performance of the protocol, we plot the spin averages in Fig. \ref{fig4}, for the various states on the Bloch sphere for different measurement outcomes $ | k \rangle $ by Alice. 
In Fig. \ref{fig4} we see that for the $ k = N $ case, nearly ideal results are obtained, where the averages of the spins agree well with the ideal outcomes. For this outcome, the closest spin expectations are obtained towards the north pole of the Bloch sphere, which originates from the fact that the 2A2S squeezed state initially starts with polarized spins at the north pole. For $ k = 0 $, the spin averages $ S^{x,y}_B $ are in good agreement with that of ideal case with the $ S^z_B $ being flipped as given in (\ref{idealspinexp}). In this case, the best parameters are for states near the south pole. On comparison with Fig. \ref{fig2}, we see that poorly performing regions near the north pole have zero probability of measurement for $ k = 0 $, hence the deviations are in fact inconsequential to the performance of the protocol.  For other values of $ k $, the spin amplitudes are diminished, as expected from the $ | 2k - N | $ factor in (\ref{idealspinexp}). There is a small non-linear contribution with respect to $ k $, where there is a different distribution to the ideal case, with a double periodicity in the $ \theta $ distribution, which can be most clearly seen for the $ k = N/2 $ case in Fig. \ref{fig4}(c). The deviations occur due to the fact that we use the 2A2S squeezed state which is not exactly the spin-EPR state.
In the above calculations we have only considered the case where the atom number on each BEC is fixed.  More realistically, when a BEC is prepared, the number of atoms changes on each cycle of the experiment.  Thus in this case the state at the beginning of the protocol of Sec. \ref{sec:protocol} is given by
\begin{align}
\rho_0 = \sum_{N_A N_B} p(N_A) p(N_B) | N_A \rangle_A \langle N_A |_A \otimes 
| N_B \rangle_B  \langle N_B |_B .  
\label{numberfluctrho}
\end{align}
Here the probabilities are considered to be Gaussian distributions 
\begin{align}
    p(N) \propto \exp ( - \frac{ (N-\bar{N})^2}{ 2\sigma_0^2 } ),
\end{align}
up to a normalization constant such that $ \sum_{N=0}^\infty p(N) = 1$.  Starting from this initial state, in Appendix \ref{app:atom number fluctuation} we derive the expression for the normalized spin average after the RSP protocol.  We calculate the normalized spin averages such that the effect of different spin amplitudes do not contribute. 

The effect of atom number fluctuations on the spin averages is shown in Fig. \ref{fig4}(e) and \ref{fig4}(f). We observe that the spin averages amplitudes are decreased and the transfer of the state has error in comparison to the ideal case as expected.  The deviation is rather small and generally remarkably robust in the presence of atom number fluctuations. We can understand this because the dependence of the protocol on the atom number fluctuations.   From the protocol shown in Sec. \ref{sec:protocol}, we can see that only Step 1 has a dependence on the number of atoms.  In Step 1, the 2A2S squeezed state must be produced at the optimal squeezing time, which has an approximate $1/N$ dependence.  For fluctuating atom number preparation, sub-optimal squeezing times will be applied to the state for a common squeezing time.  However, since the optimal time does not change very strongly within the atom number fluctuations, and furthermore becomes less sensitive for large $ N $ \cite{kitzinger2020}, this results in only a small effect on the RSP protocol.

\subsection{\label{sec:level3}Wigner Function}
The average values of the spins give only partial information about the state of the BEC on Bob’s side. To gain a better understanding of the type of state that is obtained after the RSP protocol, we calculate the spin Wigner function for various cases.  The spin Wigner function 
is a quasi-probability distribution defined as \cite{dowling1994wigner,timquantumoptics2020}
\begin{equation}
	W (\theta,\phi) = \sum_{k=0}^{2j} \sum_{q=-k}^{k} \rho_{kq} Y_{kq} (\theta,\phi),
\end{equation}
where $Y_{kq}(\theta,\phi)$ are the spherical harmonics. Here, the matrix element $\rho_{kq}$ for a given state is defined as

\begin{equation}
	\rho_{kq}=\sum_{m,m'=-j}^{j}(-1)^{j-m}\sqrt{2k+1}\begin{pmatrix}
	j&k&j\\ -m&q&m'
	\end{pmatrix}\langle jm|\rho|jm'\rangle,
\end{equation}
where $\begin{pmatrix}
j&k&j\\ -m&q&m'
\end{pmatrix}$ is the Wigner $ 3j $ symbol. This can be used to represent any state of a two component BEC with fixed particle number $ N $. Here we used a different notation for the Fock states, written in terms of angular momentum eigenstates 
\begin{equation}
	|jm\rangle=|k=j+m\rangle,
\end{equation}
where  the state on the right hand side are the Fock states as in  (\ref{fockbasis}) with $ N $ atoms.

\begin{figure}[t]%
    \includegraphics[width=\linewidth]{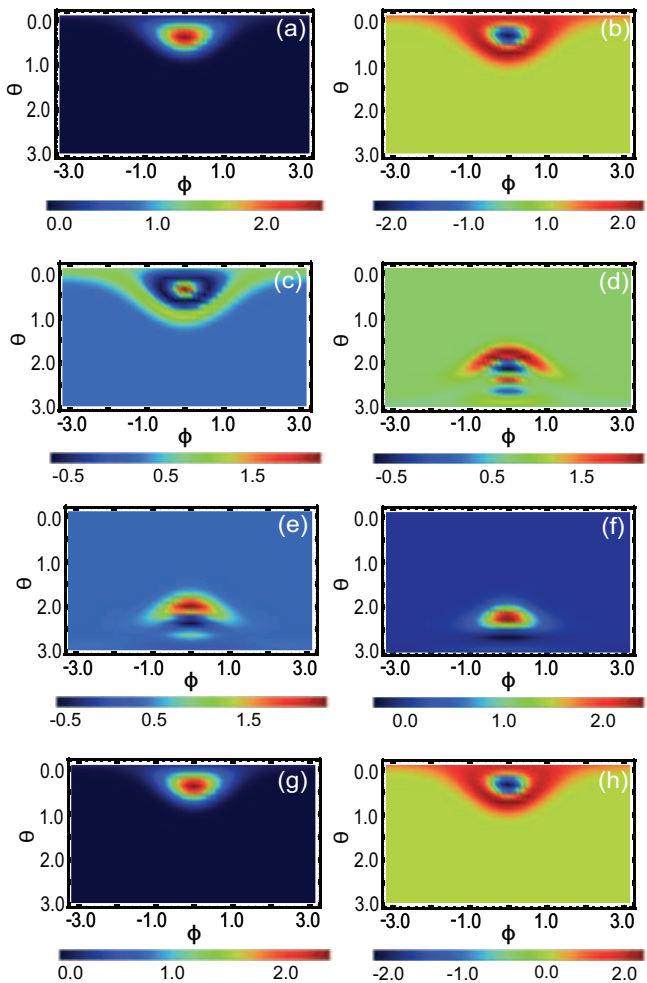}%
     \caption{Wigner functions for Bob's BEC (\ref{rspstate}) for various measurements outcomes for Alice's rotation parameters taken as $ \theta = 0.5 $ , $ \phi=0 $.  The measurement outcomes are (a) $ k = 20 $, (b) $ k = 19 $, (c) $ k = 18 $, (d) $ k = 2 $, (e) $ k = 1 $, (f) $ k = 0 $.  For comparison the ideal cases  (\ref{finalstatespinepr}) are shown for the states  (g)  $ | k = N\rangle^{(\theta, \phi)} $, (h) single Fock state $ | k = N-1\rangle^{(\theta, \phi)} $. For all cases $ N = 20 $ and       time $ \tau_{\text{opt}} = 0.1214. $     }%
    \label{fig5}%
\end{figure}

In Fig. \ref{fig5}, the Wigner functions for Bob's BEC state are plotted for different measurement outcomes by Alice labeled by $ k $ for the particular case $ \theta = 0.5$, $ \phi = 0 $. Starting with the $ k = N $ outcome in Fig. \ref{fig5}(a), we observe that the Wigner function for Bob's state is very similar to the the Wigner function corresponding to a spin coherent state \cite{timquantumoptics2020} at ($\theta$, $\phi$),  as can be seen in Fig. \ref{fig5}(g). For other non-extremal values of $ k $, such as $ k = N-1 $, the Wigner function begins to develop strong negative values, resembling the distribution of the Fock state as in (\ref{finalstatespinepr}).
In Fig.  \ref{fig5}(b) and \ref{fig5}(h) we see a comparison of the Wigner function for the outcome $ k = N - 1 $ and the state (\ref{finalstatespinepr}), which has an obvious similarity.  The case of $ k = N-2 $ corresponds to a two-particle Fock state, and the non-classical nature of the Wigner functions increase as the outcome $ k = N/2 $ is approached.  For outcomes with $ k < N/2 $, the distribution shifts to the other side of the Bloch sphere, due to the flipping of spins, as observed in (\ref{sxsyszavkminus}). Again, the non-classicality of the distributions increase towards $ k = N/2 $, and for the case $ k = 0 $, the distribution still has a remnant non-classical region due to the imperfect spin-EPR state that are being made by the 2A2S Hamiltonian.   

\begin{figure}[t]%
    \includegraphics[width=\linewidth]{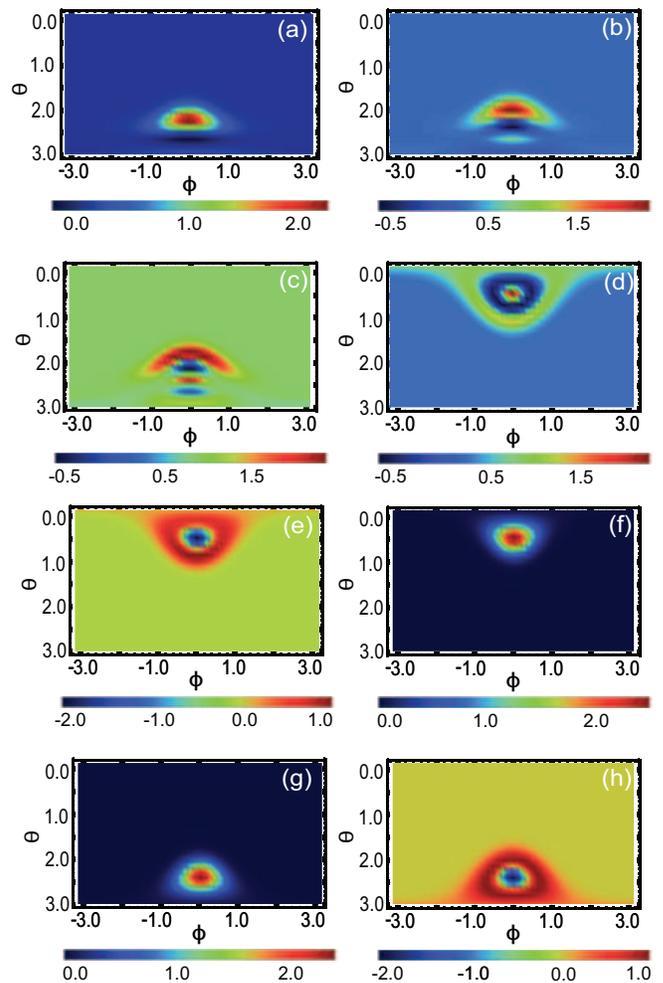}%
    \caption{Wigner functions for Bob's BEC (\ref{rspstate}) for various measurements outcomes with Alice's rotation parameters set to $ \theta = 2.5 $, $ \phi = 0 $, (a) $ k = 20 $, (b) $ k = 19 $, (c) $ k = 18 $, (d) $ k = 2 $, (e) $ k = 1 $, (f) $ k = 0 $.  For comparison the ideal cases  (\ref{finalstatespinepr}) are shown for the states  (g)  $ | k = N\rangle^{(\theta, \phi)} $, (h) single Fock state $ | k = N-1\rangle^{(\theta, \phi)} $.  For all cases $ N = 20 $ and time $ \tau_{\text{opt}} = 0.1214. $ }%
    \label{fig6}%
\end{figure}

For Alice's rotation in the southern hemisphere of the Bloch sphere $ \theta > \pi/2 $, similar behavior results, except that the relationship with the outcome $ k $ is reversed.  Fig. \ref{fig6} shows the result with $ \theta = 2.5 $, $ \phi = 0 $. Starting with the $ k = 0 $ case, we see a nearly ideal transfer of the state in terms of a spin coherent state-like Wigner distribution as seen by comparing Fig. \ref{fig6}(f) and \ref{fig6}(g), except that the distribution is reflected to the opposite hemisphere, due to the additional minus sign in $ \langle S^z \rangle $ for $ k < N/2 $. The $ k = 1 $ and $ k = 2 $ outcomes correspond to single and two particle Fock states, as seen in Fig. \ref{fig6}(e) and \ref{fig6}(d)  respectively. The non-classical nature of the distribution increases approaching $ k = N/2 $, as before. For $ k = N $ case, the distribution again has some remnant non-classicality due to the imperfect preparation of the spin-EPR state by the 2A2S squeezed state.

\subsection{Error of the remote states}
\label{sec:level4}
To measure the success of this protocol, we calculate the error of Bob's state in comparison to the ideal spin-EPR state protocol of Sec. \ref{sec:rspspinepr}. Since our aim is to prepare the Bloch sphere spin averages, one can measure the error in a similar way to the trace distance for qubits. We define this for the conditional state (\ref{rspstate}) as
\begin{align}
    E_k & (\theta,\phi) \nonumber \\
		& =\frac{1}{2N} \sqrt{\sum_{j=x,y,z}\Big(\frac{\langle \Psi_k | S^j_B | \Psi_k \rangle }{\langle \Psi_k |\Psi_k \rangle } - \langle \Psi_k^\text{ideal} | S^j_B | \Psi_k^\text{ideal} \rangle \Big)^2.
}  
     \label{overallfidelitydef}
\end{align}
Here the comparison to the ideal spin-EPR RSP protocol is given by (\ref{idealspinexp}). The expression (\ref{overallfidelitydef}) gives the distance between two states on the normalized Bloch sphere.  
In other words, it is the error of current RSP protocol measured in terms of trace distance, when it is mapped to an equivalent qubit.  The maximum possible error is 1 according to the above definition. A similar metric was used for teleporting qubit information using spin ensembles \cite{pyrkov2013,pyrkov2014full}.  
 We note that the only source of error is the imperfect preparation of the spin-EPR states, due to the use of the 2A2S squeezed state, in practice decoherence effects will potentially give further errors.  
\begin{figure}[t]%
    \includegraphics[width=\linewidth]{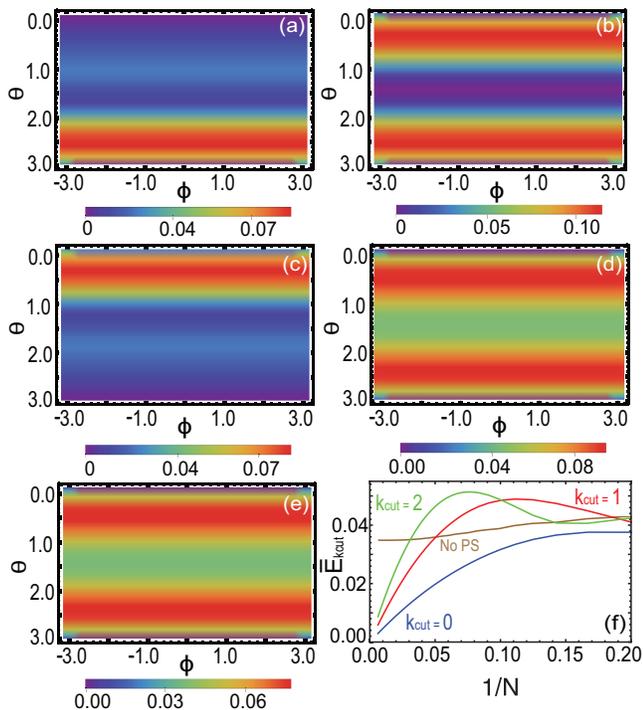}%
    \caption{Error of the RSP protocol as defined in (\ref{overallfidelitydef}) for Bob's BEC qubit for different projective measurements for $ N = 20 $, time $  \tau_{\text{opt}} = 0.1214 $ (a) $ k = N $, (b) $ k = N/2 $, (c) $ k = 0 $.  (d)(e) Average error (\ref{avg}) for $ N = 20 $, $ N =50 $ respectively.  (f) Average error for $ \theta = \pi/2 $, $ \phi = 0 $ with various $ N $ using no post-selection (\ref{avg}), labeled by ``No PS'' and post-selection (\ref{avgcutoff}), labeled by the cutoff values $ k_\text{cut} $ used.  }%
    \label{fig7}%
\end{figure}
In Fig. \ref{fig7}(a)(c) the error is plotted for Alice's measurement outcomes taking extremal 
values $k = N, 0 $ respectively.  We see that the errors are generally low except for the regions in the south and north pole of the Bloch sphere, respectively. As already observed in Fig. \ref{fig4}, the regions of high error coincide with the regions of low probability.  For example in Fig. \ref{fig2}(a) we see that for $ \theta = \pi $, the probability of $ k = N $ is zero.  Hence although it appears that high errors are achieved in some regions, it is also important to take into account the probabilities of these occuring. The reverse is true for cases $ k < N/2 $ where a high error is observed around $ \theta = 0 $ that is a low probability outcome for $ k = 0 $. For $ k = N/2 $ in Fig. \ref{fig7}(b), the best performing regions are in the region of the equator.  We point out that in this case even the ideal RSP protocol fails as the amplitude of the preparated state is zero (\ref{idealspinexp}).  Hence the origin of the errors in this case are the residual amplitudes as already observed in Fig. \ref{fig2}(c).

In order to take into account of the fact that some of the regions with poor performance coincide with low-probability events, we average the error function (\ref{overallfidelitydef}) with the probability that they occur.  We define the overall error of the RSP as
\begin{align}
    \bar{E}(\theta,\phi) = \sum_k P_k(\theta) E_k (\theta,\phi),\label{avg}
\end{align}
where $ P_k (\theta) $ is the probability of obtaining Alice's measurement outcome $ k $ as in (\ref{probability}). 
This is shown in Fig. \ref{fig7}(d)(e) for two ensemble sizes $ N $.  The best performing regions are near the two poles and in the vicinity of the equator. While the distribution of the two cases are nearly the same, the overall error tends to decrease with  $N $, as can be observed from examining the scale.   

The variation of the average error with particle number $ N $ is shown in Fig. \ref{fig7}(f) (the ``No PS'' curve). The error decreases with $ N $ as expected, but appears to approach a non-zero value. In Ref. \cite{kitzinger2020} it was  observed that the fidelity of the 2A2S squeezed state approaches a non-unit fidelity for large $ N $, although it is unclear whether logarithmic corrections are present. 
The error can be improved by performing post-selection on the measurement outcomes by Alice. We post-select results to remove the outcomes in the range $ k_{\text{cut}}< k < N - k_{\text{cut}} $. The normalized average error including post-selection is
\begin{align}
\bar{E}_{k_\text{cut}} ( \theta , \phi) = 
\frac{\sum\limits_{k\le k_{\text{cut}},k\ge N-k_{\text{cut}}} P_k ( \theta) E_k ( \theta, \phi)}{\sum\limits_{k\le k_{\text{cut}},k\ge N-k_{\text{cut}}} P_k ( \theta)}.\label{avgcutoff}
\end{align}
In Fig. \ref{fig7}(f) we show different cutoff $k_{\text{cut}}$ values. We can see that introducing post-selection improves the error particularly for larger ensembles, where the error appears to extrapolate to zero.  For $ k_{\text{cut}} = 0 $, where only Alice's outcomes  $ k = 0,N $ are kept, the error shows the best performance, at the expense of a lower success probability.  
 As we increase the $ k_{\text{cut}} $ values, a larger contribution of $ k $ values are involved including those with the poor performance, increasing the  error value.  For larger $ k_{\text{cut}} $, the curve eventually merges with the average error (\ref{avg}).

\section{\label{sec5}Summary and Conclusions}
We have introduced a RSP protocol for spin ensembles where arbitrary spin averages can be prepared using entanglement, spin rotations, and measurements in the Fock basis. This is an alternative form of the 
standard qubit RSP algorithm \cite{bennett2001remote}, that is applicable to spin ensembles.  We considered an extended version of the original qubit protocol that can prepare a state with arbitrary Bloch sphere coordinates, up to a negative sign flip on the $ S^z $.  This sign ambiguity is an existing limitation of the RSP protocol, and is the reason why typically equatorial states are only considered. The RSP protocol was constructed using only  operations and measurements that are experimentally viable, namely Fock state measurements and spin rotations. Due to the difficulty of experimentally preparing a maximally entangled spin-EPR state, we have analyzed the performance of the protocol using the 2A2S squeezed state, which serves as a close approximation to the spin-EPR state.  In particular, we examined spin averages, Wigner functions, and the error of the state that is remotely prepared in terms of the proximity with the ideal state on the Bloch sphere. We found that the transferred spin average well-approximates the ideal spin averages. The Wigner function of the transferred state is found to have close agreement with the expected state produced in the ideal spin-EPR version of the protocol. The error in terms of the distance was found to decrease with ensemble size, and the performance could be further improved using a post-selection measurement where we discard unfavorable cases.  

One of the motivations of this study is to find a simple yet non-trivial application of entanglement between two BECs. RSP is perhaps the simplest approach to entanglement-based quantum information transfer, and would be a prime candidate once entanglement between two BECs is experimentally realized.  Currently there is much  interest in entangling BECs \cite{kunkel2018spatially,lange2018entanglement,fadel2018spatial}, and we have restricted ourselves to operations that can be relatively easily performed, namely spin rotations and Fock state measurements.  The restriction on the type of measurement that we assume is the origin for the relatively simple class of states that can be produced on Bob's side.  Namely, since we assume Alice can only perform Fock state measurements, the type of state Bob receives is also a Fock state. Due to the similarity of the optimally squeezed 2A2S state to the spin-EPR state, with a more general measurement, it is likely that a wider class of states can be prepared, in a similar way to qudit RSP \cite{bennett2001remote,zeng2002remote}. We note that even with the current scheme, Fock states with highly non-classical distributions are created (stochastically) on Bob's BEC, as seen in Fig. \ref{fig5} and \ref{fig6}.  Thus one potential application of our RSP protocol is the measurement-based preparation of quantum states, which may be otherwise difficult to prepare. Another potential application is in the remote synchronization of clocks, where a similar protocol to RSP is used to transmit time information \cite{jozsa2000quantum,ilo2018remote}.

The most challenging aspect of the current protocol remains the preparation of the 2A2S entangled state.  Some options for this include first generating two-axis counterwisted squeezed states on one ensemble (i.e. 2A1S squeezed states) using methods such as that given in Ref. \cite{liu2011spin}, then performing a splitting procedure, in a similar way to Refs.  \cite{oudot2017optimal,Jing_2019}. Another way is to use optical means to generate an analogous state with similar correlations \cite{kuzmich2000generation,pettersson2017light}.  We note that our protocol differs from performing the RSP protocol in the continuous variables framework \cite{kurucz2005continuous}, since no Holstein-Primakoff approximation is used throughout our analysis.  Within the Holstein-Primakoff regime, only those states in the vicinity of the north pole would be valid in the approximation.  In contrast, for our protocol an arbitrary state on the Bloch sphere is prepared. 

In this work, we did not consider decoherence effects which will be inevitably present in a realistic experimental setting.  Since our aim is to transfer the quantum information of a single qubit, one way our protocol may be viewed is that an encoded qubit is being remotely prepared.  In this case, the encoding of the qubit is in terms of Fock states of  the appropriate basis.  A similar strategy was used in Ref. \cite{mohseni2019error} to perform an encoded version of adiabatic quantum computing, where ensembles encode an effective qubit.  In Ref. \cite{mohseni2019error} the main result was that the use of the ensembles resulted in an error suppression effect, thanks to the duplication of the quantum information.  A simple way to understand the error suppression effect of using ensembles is that it results in a boosted signal-to-noise. For example, if the  typical amplitude of the spins is $ \sim N $, and if a  depolarizing channel acts on the spin, this would modify the spin expectation values to $\sim \epsilon N$.  The boost of $N$ provides a much larger signal to work with, in comparison to single qubits $ N = 1 $. Since in BECs, $ N $ can be $10^3$  \cite{riedel2010atom,fadel2018spatial}, and even larger for atomic ensembles \cite{julsgaard2001experimental}, this provides a considerable boost. We can thus anticipate that the use of spin ensembles should provide an error suppression effect in a similar way to Ref. \cite{mohseni2019error}.  A more detailed investigation of this will be left as future work.  

\begin{acknowledgments}
This work is supported by the Shanghai Research Challenge Fund; New York University Global Seed Grants for Collaborative Research; National Natural Science Foundation of China (61571301,D1210036A); the NSFC Research Fund for International Young Scientists (11650110425,11850410426); NYU-ECNU Institute of Physics at NYU Shanghai; the Science and Technology Commission of Shanghai Municipality (17ZR1443600,19XD1423000); the China Science and Technology Exchange Center (NGA-16-001); and the NSFC-RFBR Collaborative grant (81811530112). A.N.P acknowledges the RFBR-NSFC collaborative program (Grant No. 18-57-53007) and the State assignment (N. 0089-2020-0002).
\end{acknowledgments}

\appendix
\section{Expression for transformation of Fock states in various spin bases}
\label{app:prob}
The Fock states $|k\rangle $ in (\ref{fockbasis}) are eigenstates of the $ S^z $ spin operator, one can transform it to an arbitrary direction $| k \rangle^{(\theta, \phi)}$ in Ref. \cite{timquantumoptics2020}.
The rotation operations in the $ S^y $ and $ S^z $ directions consecutively transforms the Fock state as
\begin{align}
   &|k\rangle^{(\theta, \phi)} =  e^{-iS^z \phi/2} e^{-iS^y \theta/2} | k \rangle =  \sum_{k'}e^{-i(2k'-N)\frac{\phi}{2}} \nonumber \\ &  \times  \sqrt{k!(N-k)!k'!(N-k')!}  \nonumber \\ &   \times  \sum_{n=\text{max}(k-k',0)}^{\text{min}(k,N-k')}\frac{(-1)^n}{(k'-n)!(N-k-n)!n!(k-k'-n)!} \nonumber \\
 & \times   \cos^{k'-k+N-2n}(\theta/2)\sin^{2n+k-k'}(\theta/2) |k'\rangle .
      \label{matrixelem}
\end{align}

\section{Effect of number fluctuations}
\label{app:atom number fluctuation}
In a realistic BEC, the number of atoms fluctuates from shot-to-shot due to the probabilistic processes involved in its preparation.  In this section, we show the effect of number fluctuations on the RSP protocol.  

The ensemble average of the initial state taken over many shots at the start of the protocol according to (\ref{numberfluctrho}).  
Starting from this initial state, proceeding with the steps shown in Sec. \ref{sec:protocol} gives the normalized state
\begin{align}
\rho(\tau) =  \sum_{N_A N_B} p(N_A) p(N_B) \frac{ |  \Psi_k^{(N_A, N_B)} \rangle  \langle \Psi_k^{(N_A, N_B)}   |  }{ \langle \Psi_k^{(N_A, N_B)} |  \Psi_k^{(N_A, N_B)} \rangle  }    
\end{align}
where we defined
\begin{align}
| \Psi_k^{(N_A, N_B) }  (\tau) \rangle = & U_k^C | k \rangle \langle k |_A  {U^{(\theta, \pi - \phi )}}^\dagger e^{iS^z_A \pi/ 8 }  e^{iS^z_B \pi/ 8 } \nonumber \\
& \times e^{-i(S_A^+ S_B^+ + S_A^- S_B^-) \tau }
| N_A \rangle_A | N_B \rangle_B .  
\end{align}
This explicitly labels the particle numbers of each BEC in the wavefunction (\ref{rspstate}).
The spin averages are then evaluated for the number fluctuating case as
\begin{align}
\langle s^j_B \rangle = & \sum_{N_A N_B} \frac{p(N_A) p(N_B)}{N_B} \nonumber \\
& \times \frac{ \langle \Psi_k^{(N_A, N_B)} (\tau)  | S^j_B | \Psi_k^{(N_A, N_B)} (\tau) \rangle    }{ \langle \Psi_k^{(N_A, N_B)} (\tau) |  \Psi_k^{(N_A, N_B)} (\tau) \rangle  }   . \label{stat.}
\end{align}

\bibliography{rsp}


\end{document}